\documentstyle[12pt]{article}
 
\begin{document}
\author{\and J.~Goedert and F.~Haas \\
Instituto de F\'{\i}sica, UFRGS\\
Caixa Postal 15051\\
91500-970 Porto Alegre, RS - Brazil}
\title{On the Lie Symmetries of a Class of Generalized Ermakov Systems}
\date{\relax}
\maketitle

\begin{abstract}
The symmetry analysis of Ermakov systems is extended to 
the generalized case where the frequency depends on the 
dynamical variables besides time. In this extended framework, a 
whole class of nonlinearly coupled oscillators are viewed as
Hamiltonian Ermakov system and exactly solved in closed form. 

\end{abstract} 

\section{Introduction}

The first study of an Ermakov system was published by the end of 
the last century \cite{Ermakov}, but the motivation for a
detailed investigation of Ermakov systems intrinsic properties, 
their proper characterization and their application scope derived
mainly from the works of Lewis who rediscovered them in 
the sixties  \cite{Lewis1}, and of Lewis and Riesenfeld, who 
devised one of their most important application \cite{Lewis2}, 
the quantization of the time-dependent harmonic oscilator.  

In its original formulation, the Ermakov system was nothing but a 
time-dependent harmonic oscillator coupled to Pinney's 
\cite{Pinney} equation. Later on, Ray and Reid \cite{Ray1} extended 
the basic concept by incorporating two extra arbitrary 
functions in the original equations. Such generalization highly 
improved the reach of the Ermakov systems with no harm to their key 
property, namely the existence of a constant of motion, the 
Ermakov--Lewis invariant.  

Early developments dealt with Ermakov systems where the 
frequency \cite{Ray1} depended solely on time. In a subsequent 
generalization \cite{Reid}, the dynamical variables were 
allowed to participate in the argument of the frequency.  
The resulting generalized systems have rather interesting 
properties and potential applications \cite{Goedert, Haas}, but has 
stayed mostly as a fertile field for further prospects. In particular, the 
Lie point symmetry analysis of Ermakov systems  performed by Leach 
and co-workers \cite{Leach}--\cite{Govinder3} considered frequency 
functions depending at most on time. Inspired by the  
results of Leach and co-workers, we reconsider the issue of 
geometric symmetries for {\it generalized} Ermakov systems
where the frequency is a function of both the dynamical 
variables and their first derivatives besides time.  By doing this 
we  find a much wider class of dynamical systems that aside from 
possessing the Ermakov--Lewis invariant admits a Lie symmetry in
a direct and natural generalization of that proposed 
by Leach.  As a first usefull application of this extended 
viewpoint we show that some systems which would otherwise not fit 
in the older framework are readily treated as Ermakov systems.   

This letter is organized as follows. In section II, the class of 
generalized  Ermakov systems with Lie point symmetry is 
determined.  In Section III a class of  coupled nonlinear oscillators
are shown to constitute a generalized Ermakov system that possesses
geometric symmetries and are exactly solvable in closed form. 
The conclusions are presented in section V.  

\section{The Symmetry Analysis}

All Ermakov systems, both usual and generalized, 
can be represented \cite{Haas} by the pair of equations
\begin{eqnarray}
\label{erma1}
\ddot x + \Omega^2 x &=& \frac{1}{yx^2} F(y/x) , \\
\label{erma2}
\ddot y + \Omega^2 y &=& 0 ,
\end{eqnarray}
where $F$ is an arbitrary function of the indicated argument and 
$\Omega$, hereafter called the generalized frequency, is an 
arbitrary function of the time, the coordinates and their derivatives.  
For physical reasons, we restrict the study to the cases where the 
frequency depends at most on the velocities, that is, we take 
$\Omega = \Omega(t,x,y,\dot{x},\dot{y})$ only.  

The pair of equations (\ref{erma1}--\ref{erma2})
encompasses the whole class of Ermakov systems in two spatial 
dimensions. In the traditional notation, the Ermakov systems were 
written in terms of {\it three} arbitrary funtions \cite{Ray1} and 
not only {\it two} as we have proposed \cite{Haas} elsewhere. This 
more concise representation is possible thanks to the generalized 
character of $\Omega$.  The usual representation is obtained 
whenever $\Omega$ is chosen in the form 
\begin{equation}
\label{Omega}
\Omega^2 = \omega^{2}(t) - \frac{1}{xy^3}g(x/y) \,,
\end{equation}
where $g(x/y)$  and $\omega$ are arbitrary functions of the 
indicated arguments.  

The use of generalized frequencies considerably expands the reach 
of the Ermakov system concept. Among the systems that can be 
treated as particular Ermakov systems in the generalized form, we 
quote the Kepler-Ermakov system \cite{Athorne1} and the Lutzky's 
integrable system \cite{Lutzky}.  

In order to perform the Lie symmetry analysis of generalized 
Ermakov system (\ref{erma1}--\ref{erma2}) we follow the approach of 
Govinder and Leach for the usual case \cite{Leach, Govinder3}, 
and consider initially the symmetries of the  equivalent system 
\begin{eqnarray}
\label{sim1}
x \ddot y - y \ddot x + 
\frac{1}{x^2} F(y/x) = 0 \,, \\
\label{sim2}
\ddot y + 
\Omega^{2}({\bf x}, {\bf\dot x}, t) y = 0 \,, 
\end{eqnarray}
where the notation ${\bf x} = (x,y)$ was introduced. The advantage 
of dealing first with the equivalent equation (\ref{sim1}) derives 
from its independence of $\Omega$. This feature  provides a 
straightforward way of finding its symmetry generator, which is 
basically that  obtained by Govinder and Leach \cite{Govinder3}
in the symmetry analysis of the usual Ermakov systems, 
\begin{equation}
\label{gera}
G = \rho^{2}\frac{\partial}{\partial t} + 
\rho\,\dot\rho\,{\bf x}\cdot\frac{\partial}{\partial{\bf x}} 
\,,
\end{equation}
where $\rho(t)$ is an arbitrary differentiable function.  As 
pointed out in ref. \cite{Govinder3}, the Ermakov first integral $I$ 
is invariant under the first extension of $G$.  

For the complete Ermakov system, that is, for equations 
(\ref{sim1}--\ref{sim2}) to obey the same symmetry, some 
restriction must be imposed on the functional form of $\Omega$. 
This restriction is determined by requiring the invariance of Eq. 
(\ref{sim2}) under the action of the first extension of $G$. We
apply this criterion, and obtain the most general frequency that 
preserves the invariance of the generalized Ermakov system under 
the action of $G$, 
\begin{equation}
\label{allowable}
\Omega^2 = - \frac{\ddot\rho}{\rho} + 
\frac{1}{\rho^4} \sigma({\bf x}/\rho, \rho\dot{\bf x}-\dot\rho{\bf x}) \,, 
\end{equation}
where $\sigma$ is an arbitrary function of the indicated arguments.  
Notice that velocity dependence is also possible, a fact that 
considerably expands the class of Ermakov systems invariant under 
the symmetry transformations of $G$.  

To summarize the results so far we note that with no {\it a  
priori} assumption on the functional form of the frequency, we find 
that the  most general Ermakov system possessing Lie 
point symmetry must be of the form 
\begin{eqnarray}
\label{erin1}
\ddot x + \left(\frac{\sigma}{\rho^4} - \frac{\ddot\rho}{\rho}\right) x &=& 
\frac{1}{yx^2} F(y/x) \,, \\
\label{erin2}
\ddot y + \left(\frac{\sigma}{\rho^4} - \frac{\ddot\rho}{\rho}\right) y &=& 0\,.
\end{eqnarray}
 where $\sigma$ is arbitrary but restricted to
\begin{equation}
\label{sigmaf}
\sigma=\sigma({\bf x}/\rho, \rho\dot{\bf x}- \dot\rho{\bf x}) \,.
\end{equation}
The associated transformation group is that generated by $G$.

Let us show now that the whole class of usual Ermakov 
systems are recovered as special cases of the 
present treatment. For this purpose we use the relation 
\begin{equation} 
\label{pinup} 
\ddot\rho + \omega^{2}(t)\rho = \frac{\Omega_{0}^2}{\rho^3} \,, 
\end{equation}
to {\it define} a new function $\omega(t)$ in terms of $\rho$
and the constant, but otherwise arbitrary, $\Omega_{0}$.  
Despite its appearance of a differential equation, 
(\ref{pinup}) can be alternatively considered as the definition 
of $\omega$ in terms of $\rho$. Let us adopt this viewpoint and 
also define a transformed $\bar{\sigma}$ 
\begin{equation}
\bar{\sigma}({\bf x}/\rho, \rho\dot{\bf x} - 
\dot\rho{\bf x}) = - \Omega_{0}^2 + \frac{\rho^4}{xy^3} g(x/y) + \sigma \,,
\end{equation} 
where we introduced, for convenience, a spurious but otherwise 
innocuous function $g(x/y)$, which can be absorbed in the 
definition of $\sigma$.  These redefinitions transform the generalized 
form of the Ermakov system (\ref{erin1}--\ref{erin2}) into its more 
traditional form \cite{Ray1} 
\begin{eqnarray}
\label{equa1}
\ddot x + (\omega^{2}(t) + \frac{\bar{\sigma}}{\rho^4}) x 
             &=& \frac{1}{yx^2} f(y/x) \,, \\
\label{equa2}
\ddot y + (\omega^{2}(t) + \frac{\bar{\sigma}}{\rho^4}) y 
            &=& \frac{1}{xy^2} g(x/y) \,.
\end{eqnarray}
In equations (\ref{equa1}--\ref{equa2}) we replaced
\begin{equation}
 F(y/x)= f(y/x) + \frac{x^2}{y^2} g(x/y) 
\end{equation}
in order to facilitate the comparison with previous results 
on the Lie group of usual Ermakov systems. We should stress, 
however, that the superfluous  function $f(y/x)$ was introduced 
only in order to reconstruct the traditional form.  

We now notice that the pair of equations (\ref{equa1}--\ref{equa2}) 
reduces to the usual form with simple time dependence in the 
frequency only when $\bar{\sigma} \equiv 0$.  For this particular
class of frequencies the quasi-invariance transformation 
\cite{Burgan}  
\begin{equation} 
\label{quasi} \bar{x} = x/C \qquad \bar{y} = y/C 
\qquad \bar{t} = \int\,dt/C^2 \,, 
\end{equation} 
removes $\omega$ \cite{Athorne2} from the equations of motion,
which become 
\begin{equation} 
\label{sl2} 
\bar{x}'' = \frac{1}{\bar{y}\bar{x}^2} 
f(\bar{y}/\bar{x}) \,, \qquad \qquad
\bar{y}'' = \frac{1}{\bar{x}\bar{y}^2} 
g(\bar{x}/\bar{y}) \,,  
\end{equation} 
whenever $C(t)$ is a solution of the time-dependent harmonic oscillator 
\begin{equation} 
\label{tdho} \ddot C  + \omega^{2}(t)C = 0 \,.
\end{equation}
The symmetry group of the transformed Ermakov system (\ref{sl2}) is  
$SL(2,R)$ as already shown in \cite{Leach}.  This however is not 
the case with equations (\ref{equa1}--\ref{equa2}) which are more 
general than the  Ermakov system with $SL(2,R)$ symmetry.  The extra 
symmetry freedom of the system (\ref{equa1}--\ref{equa2}) derives 
from $\bar{\sigma}$, which can depend on the coordinates or the 
velocities. Notice however that the transformations belonging to 
the $SL(2,R)$ group are among the transformations generated by $G$.  

The symmetries considered so far do not require any particular form 
for the function $\sigma$, nor depend on any preliminary coordinate 
transformation like (\ref{quasi}).  In this respect, it should be 
noted that, despite its apparent simplicity, the transformation 
(\ref{quasi}) is not  well defined when $\omega$ depends on the 
dynamical variables, a situation in which equation (\ref{tdho}) 
becomes meaningless.  In addition, even for $\omega = \omega(t)$, 
the time-dependent harmonic oscillator (\ref{tdho}) does not always 
possess a closed form solution, a fact that may preclude its 
application in many interesting practical situation.  Similar 
remarks apply to the linearization of the Ermakov system proposed in 
\cite{Athorne2}, which starts exactly with the transformation 
(\ref{quasi}). Needless to say, such a linearization process 
applies only to the simplified system (\ref{sl2}) and not to the
generalized Ermakov system (\ref{erin1}--\ref{erin2}). However, it 
should be stressed that other generalized Ermakov systems involving 
extensions to multi-components and higher dimensions have recently 
been introduced  which do have the underlying linear structure of 
the Ray-Reid system \cite{rogersp, bassom}. 

\section{Hamiltonian Generalized Ermakov Systems} 

The Ermakov systems with $SL(2,R)$ symmetry in two spatial dimensions 
are solvable in closed form \cite{Govinder2}.  The more general 
Ermakov systems (\ref{erin1}--\ref{erin2}), however, are not always 
exactly solvable and this, certainly, is the price paid for the 
extra generality. Some additional structure, perhaps a Hamiltonian 
formalism, must exist in order to reobtain equations that allow for 
exact solutions. To present one application example and to gain 
insight in this respect, let us consider  the two dimensional 
oscillator with nonlinear coupling introduced by Ray and Reid 
\cite{Ray2} 
\begin{eqnarray}
\label{osc1}
\ddot x + \omega^{2}(t) x &=& x\rho^{-4} G(xy/\rho^2) \,, \\
\label{osc2}
\ddot y + \omega^{2}(t) y &=& y\rho^{-4} G(xy/\rho^2) \,, 
\end{eqnarray}
where $\rho(t)$ is an auxiliary function satisfying Pinney's equation 
\begin{equation}
\label{pinn}
\ddot\rho + \omega^{2}(t)\rho = 1/\rho^{3} \,.
\end{equation}
Ray and Reid found a first integral for (\ref{osc1}--\ref{osc2}), namely
\begin{equation}
\label{constant}
J = (\rho\dot x - \dot\rho x) (\rho\dot y - \dot\rho y) + 
\frac{xy}{\rho^2} - \int^{xy/\rho^2} G(\tau)\,d\tau \,,
\end{equation}
but did not obtain its complete solution.  In this section we show 
that the generalized Ermakov system approach combined with the 
symmetry analysis, yields the corresponding complete solution, in 
closed form.  

Let us first show that (\ref{osc1}--\ref{osc2}) is in fact a 
generalized Ermakov system.  This is easily verified by choosing
in (\ref{erin1}--\ref{erin2}), $F = 0$ and $\Omega$ a function of 
the coordinates
\begin{equation}
\label{rayomega}
\Omega^2 = \omega^{2}(t) - 
\frac{1}{\rho^4} G(xy/\rho^2) \,.
\end{equation}
From this viewpoint the amplitude of the angular momentum
\begin{equation}
\label{rayermakov}
I = \frac{1}{2}(x\dot y - y\dot x)^{2}\,,
\end{equation}
becomes the corresponding Ermakov--Lewis invariant.  Moreover, 
(\ref{osc1}--\ref{osc2}) form a Hamiltonian Ermakov system, a  
property that has already played a major role in several 
applications \cite{Haas, Cervero, Athorne3}.  For the present case 
the Hamiltonian is 
\begin{equation}
H = p_{x}p_{y} + \omega^{2} xy - 
\frac{1}{\rho^2} \int^{xy/\rho^2} G(\tau)\,d\tau \,,
\end{equation}
which is not a constant of motion, since the system is non-conservative.  

The frequency (\ref{rayomega}) is of the right form (\ref{allowable}) 
and therefore guaranties the invariance of (\ref{osc1}--\ref{osc2}). The
use of the auxiliary equation (\ref{pinn}) to substitute $\rho$ for 
$\omega(t)$, transforms $\Omega$ into 
\begin{equation}
\Omega^2 = - \frac{\ddot\rho}{\rho} + \frac{1}{\rho^4} 
(1 - G(xy/\rho^2)) \,, 
\end{equation}
which is indeed of the general form (\ref{allowable}).   

The Hamiltonian structure and the symmetry can now be further 
exploited to reduce the problem to quadrature.  For this purpose, 
the use of canonical group coordinates,   
\begin{equation}
\label{transs}
u = x/\rho \,, \qquad
v = y/\rho \,, \qquad 
T = \int \rho^{-2} \,dt \,,
\end{equation}
is an essential artifice that transforms the symmetry generator 
into a simple time translation, $G = \partial/\partial T$.  In this 
new variables, the equations of motion become the autonomous system 
\begin{equation}
u'' + u = u G(uv) \,, \qquad\qquad v'' + v = v G(uv) \,, 
\end{equation}
where primes denote derivatives with respect to the new 
time. Moreover, this system is described by the Hamiltonian
\begin{equation}
\label{ka}
K = p_{u}p_{v} + uv - \int^{uv} G(\tau) \,d\tau \,,
\end{equation}
which is the transformed version of the first integral $J$ given by 
(\ref{constant}). The Hamiltonian $K$ in the canonical group 
coordinates is quadratic in momentum. In such cases \cite{Haas}, the 
appropriated variables for quadrature are 
\begin{equation}
q = uv \,, \quad 
s = v/u \,.
\end{equation}
In these coordinates, the constants of motion become
\begin{eqnarray}
\label{inte1}
\sqrt{2I} &=& q\,s'/s , \\
\label{inte2}
J &=&  (q'^{2} - 2I)/4q + q - \int^{q} G(\tau)\,d\tau \,.
\end{eqnarray}
The quadrature of the last two equations give successively $q(T)$ 
and $s(T)$.  The map back to the original variables is a pure 
algebraic task of using 
\begin{equation}
x^2 = \rho^{2}q/s \,, \qquad
y^2 = \rho^{2}q\,s \,, 
\end{equation}
to  substitute $(x,y)$ for $(q,s)$.  One final step is the 
determination of $T(t)$ which can be obtained by the integration of 
the last equation in (\ref{transs}).  In the quadrature sense, this 
completes the integration process.  Recall that all the necessary 
structure and the correct choice of variables  was dictated by the 
Hamiltonian character, the generalized Ermakov form of the system 
and by the presence of symmetry.  In particular, the analytical 
form of the solution can be found explicitly in terms of elliptic 
functions when $G(\tau) = c_{1}/\tau^2 + c_{2} + c_{3}\tau + 
c_{4}\tau^2$, where $c_i$ are arbitrary constants.  %

\section{Conclusion}

Generalized Ermakov systems (\ref{erin1}--\ref{erin2}) are of a 
very general nature, and their complete application scope is 
still not fully outlined. In this paper we determined their 
Lie point symmetry group and showed that a set of nonlinear 
coupled oscillators belonging to this class of systems 
is in fact explicitly soluble in term of quadratures. This provides 
a stimulating preview of the variety of applications in 
which generalized Ermakov systems may play an important role.  

To conclude, we mention that Lie symmetries of the type 
generated by (\ref{gera}) and some modifications of it
occur in several problems in Physics.  Some non Hamiltonian
systems with velocity dependent frequency, a whole class of two and 
three-dimensional charged particle motion, some perturbed magnetic 
monopole and a time-dependent Kepler problem fall in the category of 
such systems \cite{work} and are frequently amenable to exact 
solution in closed form.  

\section*{Acknowledgments}
This work has been partially supported by the Brazilian agencies 
Conselho Nacional de Desenvolvimento Cient\'{\i}fico e 
Tecnol\'ogico~(CNPq) and Financiadora de Estudos e 
Projetos~(FINEP).  

\bigskip

\end{document}